\documentclass{aa}  

\usepackage{natbib}
\usepackage{savesym}
\usepackage{euscript}
\usepackage{graphicx,bm,times,url}
\usepackage{amsmath,mathtools}
\usepackage[varg]{txfonts}
\usepackage[colorlinks, citecolor=blue]{hyperref}
\usepackage[normalem]{ulem}
\usepackage{xcolor}
\graphicspath{{./fig/}{./png/}}

\newcommand{\ee}{\mbox{\boldmath $e$} {}}

\newcommand{\m}[1]{\mathcal{#1}}
\newcommand{\healpix}{{\tt HEALPix}}
\newcommand{\planck}{{\it Planck}}

\newcommand{\lan}{\langle}
\newcommand{\ran}{\rangle}
\newcommand{\E}{$E$}
\newcommand{\B}{$B$}
\newcommand{\TEB}{$T$-\E-\B}
\newcommand{\I}{$I$}
\newcommand{\Q}{$Q$}
\newcommand{\U}{$U$}

\begin{document}

   \title{The link between \E-\B\ polarization modes and gas column density from interstellar dust emission\thanks{Based on observations obtained with Planck (\url{http://www.esa.int/Planck}), an ESA science mission with instruments and contributions directly funded by ESA Member States, NASA, and Canada.}}

   \author{Andrea Bracco
          \inst{1} \and
          Tuhin Ghosh 
          \inst{2} \and
          Francois Boulanger 
          \inst{1,3}\and
          Jonathan Aumont 
          \inst{4}
          }

   \institute{Laboratoire de Physique de l’Ecole Normale Supérieure, ENS, Université PSL, CNRS, Sorbonne Université, Université de Paris, Paris, France \\ 
              \email{andrea.bracco@ens.fr}
         \and
              School of Physical Sciences, National Institute of Science Education
and Research, HBNI, Jatni 752050, Odisha, India
         \and
             Institut d'Astrophysique Spatiale, CNRS (UMR8617) Universit\'e Paris-Sud 11, B\^atiment 121, Orsay, France
          \and 
          IRAP, Universit\'e de Toulouse, CNRS, CNES, UPS, (Toulouse), France
             }

   \date{Received: 24 May 2019 ; Accepted: XXX}

 
  \abstract
  {The analysis of the \planck\ polarization \E\ and \B\ mode power spectra of interstellar dust emission at 353 GHz recently raised new questions on the impact of Galactic foregrounds to the detection of the polarization of the Cosmic Microwave Background (CMB) and on the physical properties of the interstellar medium (ISM). In the diffuse ISM at high-latitude a clear $E-B$ asymmetry is observed, with twice as much power in \E\ modes than in \B\ modes; as well as a positive correlation between the total power, $T$, and both \E\ and \B\ modes, presently interpreted in terms of the link between the structure of interstellar matter and that of the Galactic magnetic field.}
{In this paper we aim at extending the \planck\ analysis of the high-latitude sky to low Galactic latitude, investigating the correlation between the \TEB\ auto- and cross-correlation power spectra with the gas column density from the diffuse ISM to molecular clouds.}
{We divide the sky between Galactic latitude $|b|>5^{\circ}$ and $|b|<60^{\circ}$ in 552 circular patches, with an area of $\sim$400$^\circ{^2}$, and we study the cross-correlations between the \TEB\ power spectra and the column density of each patch using the latest release of the \planck\ polarization data.}    
{We find that the $B$-to-$E$ power ratio ($\m{D}^{BB}_{\ell}/\m{D}_{\ell}^{EE}$) and the $TE$ correlation ratio ($r^{TE}$) depend on column density. While the former increases going from the diffuse ISM to molecular clouds in the Gould Belt, the latter decreases. This systematic variation must be related to actual changes in ISM properties. The data show significant scatter about this mean trend. The variations of $\m{D}^{BB}_{\ell}/\m{D}_{\ell}^{EE}$ and $r^{TE}$ are observed to be anti-correlated for all column densities. In the diffuse ISM, the variance of these two ratios is consistent with a stochastic non-Gaussian model in which the values of $\m{D}^{BB}_{\ell}/\m{D}_{\ell}^{EE}$ and $r^{TE}$ are fixed. We finally discuss the dependencies of $TB$ and $EB$ with column density, which are however hampered by instrumental noise.     
}
{For the first time, this work shows significant variations of the \TEB\ power spectra of dust polarized emission across a large fraction of the Galaxy. Their dependence on multipole and gas column density is key for accurate forecasts of next generation CMB experiments and for constraining present models of ISM physics (i.e., dust properties and interstellar turbulence) that are considered responsible for the observed \TEB\ signals.}

   \authorrunning{A. Bracco et al.}
   \titlerunning{\E-\B\ modes of dust polarization versus gas column density}
   \maketitle
%
\begin{figure*}[ht] 
   \centering
   \includegraphics[width=0.9\textwidth]{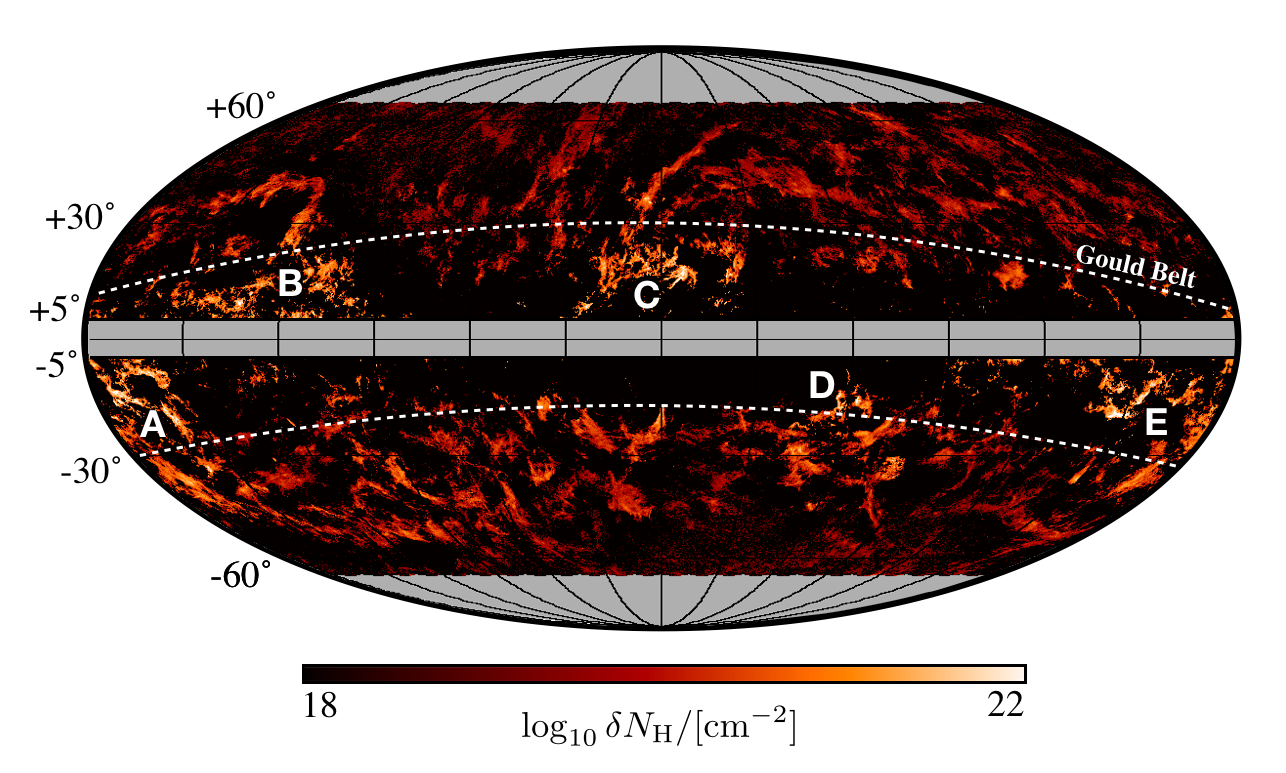}
   \caption{Filtered column density map, $\delta N_{\rm H}$, derived from {\it Planck} data of interstellar dust emission. The Gould Belt is qualitatively represented by the two white-dashed lines. Within the Gould Belt  some bright and close-by molecular clouds can be identified (< 600 pc from the Sun): Taurus, Perseus, and California in the extreme East (A); Cepheus and Polaris in the North-East (B); Ophiuchus above the Galactic center (C); Musca and Chamaeleon in the South-West (D); Orion in the extreme West (E).  A Galactic coordinates grid centered in ($l,b$)=(0$^\circ$,0$^\circ$) is added with steps of $30^\circ$ both in longitude and in latitude. 
   }
   \label{fig:dNH}
\end{figure*}
\section{Introduction}\label{intro}
The Galactic polarized light emitted by interstellar dust grains is considered a major foreground for detecting primordial \B-modes of the Cosmic Microwave Background (CMB) \citep[][hereafter P16XXX]{planck2014-XXX}. The \E-\B\ mode decomposition was introduced by \citet{Zaldarriaga1997} to characterize the polarization of the CMB as it allows one to build an orthogonal base for linear polarization that is invariant under rotation, in contrast to the Stokes parameters \Q\ and {\U}, and separates the CMB polarization in components of different physical origins. More generally, as \E\ and \B\ modes are scalar (parity-even) and pseudo-scalar (parity-odd) quantities, respectively, their auto- and cross-correlation power spectra are ideal to probe the two-point statistics in polarization across the sky. 

In the case of CMB, the \B\ mode power would partly be the result of tensor perturbations in the early Universe generated by primordial gravitational waves during the epoch of cosmic inflation \citep{Kamionkowski1997}. Such a detection would represent an indirect proof of the paradigm of cosmic inflation after the Big Bang. Until now a wealth of experiments from the ground (e.g., DASI \citep{Carlstrom2000}, ACT \citep{Marriage2009}, POLARBEAR \citep{Kermish2012}, BICEP1/2 \citep{Pryke2013}), balloon (e.g., BOOMERanG \citep{DeBernardis2000},
SPIDER \citep{Fraisse2013}), and satellite (e.g., WMAP \citep{Bennett2013}, {\planck} \citep{PlanckI2016}), have reached the required sensitivity to perform accurate measurements of the CMB anisotropies both in intensity and in polarization. However, the extraction of the cosmological signal is still limited by the ability of controlling instrumental systematics and subtracting foreground contamination that
add to the primordial radiation. 

Above $100$~GHz the most important CMB foreground is interstellar dust emission. Thanks to the first full-sky maps in polarization at 353 GHz obtained with the {\it Planck} satellite \citep{PIPXIX2015}, it has been possible to quantify the levels of \E\ and \B\ modes from Galactic dust. Focusing on the high/intermediate-Galactic-latitude sky ($|b|>35^\circ$) P16XXX, first, and more recently \citet[][hereafter P18XI]{PIPXI2018}, showed that on average 
(i) the dusty Milky Way produces twice as much power in \E\ modes than in the \B-modes (also referred to as $E-B$ asymmetry); (ii) a positive correlation exists over a wide range of angular scales (for multipoles $\ell > 5$) between \E-modes and the total intensity, Stokes \I, alternatively referred to as $T$; (iii) a hint of a positive correlation at large angular scales (for multipoles $\ell < 100$) between $T$ and \B-modes is present as well.

The origin of these observational results is yet to be established. More work is therefore needed to model them as CMB foregrounds. They are the consequence of the physical processes in the interstellar medium (ISM) that generate and affect dust polarization. Dust grains aligned with the interstellar magnetic field \citep[i.e., ][]{Chandra1953, Davis1951, Lazarian2007, HoangLazarian2016, HoangLazarian2018} and mixed with interstellar gas emit thermal radiation with a polarization vector preferentially perpendicular to the local orientation of the  magnetic field. Hence, dust polarization observations are a suitable probe of the physical coupling between the gas dynamics and the magnetic-field structure, giving insight into magnetohydrodynamical (MHD) turbulence in the ISM \citep[e.g.,][]{Brandenburg2013}. 

The possibility that the cross-correlations between dust polarization power spectra are related to MHD turbulence in the ISM has been recently investigated by several authors, although no general agreement has been achieved yet. \citet{Kritsuk2017} and \citet{Kandel2017,Kandel2018} suggested that sub-Alfv\'enic turbulence at high-Galactic latitude (with Alfv\'en Mach number $M_{A}<0.5$) may reproduce the \E-\B\ asymmetry and the positive $T$-\E\ correlation at $\ell > 50$. \citet{Caldwell2017} on the contrary concluded that only a narrow range of theoretical parameters in MHD simulations would account for the observations, suggesting that \planck\ results may likely connect to the large-scale driving of ISM turbulence. The \E-\B\ asymmetry was also found to be produced by inhomogeneous helical turbulence in \citet{Brandenburg2019}, investigating the role of magnetic helicity in the emergence of parity-odd/even quantities in interstellar polarized emission. The variety and complexity of simulated scenarios able to reproduce the \E-\B\ decomposition from \planck\ is described as well in \citet{Kim2019}. The authors presented a first statistical analysis of all-sky synthetic maps of dust polarization at $353$~GHz produced with the TIGRESS MHD simulations. Displacing the view point within a kpc-scale shearing box, they found large 
fluctuations of \E-\B\ asymmetry and $T$-\E\ correlation depending both on the observer's position and on temporal fluctuations of ISM properties due to bursts of star formation.   
The observer's environment, and the role of the large-scale Galactic magnetic field in the Solar neighborhood, were also considered in \citet{Bracco2019} as a possible explanation for the positive $T$-\E\ and $T$-\B\ correlations at very low multipoles ($\ell < 50$) via a left-handed helical component.

For multipoles $\ell > 50$, sub/trans-Alfv\'enic turbulence in the diffuse ISM was independently suggested by additional observational evidence. Sub/trans-Alfv\'enic turbulence would explain the overall alignment of the magnetic-field morphology with the distribution of filamentary matter-density structures observed with dust emission at high Galactic latitude \citep{PIPXXXII,planck2015-XXXVIII,Soler2017}. The alignment between density structures and magnetic fields, as suggested by \citet[][hereafter Z01]{Zaldarriaga2001}, would generate more $E$-mode power compared to the $B$-modes and naturally explain the positive correlation between $T$ and $E$, at least on angular scales typical of interstellar filaments (for multipoles $\ell > 50$). 

The analysis of the histograms of relative orientation (HROs) between magnetic-field and density structures showed a change in trend from the diffuse ISM to dense molecular clouds in the Galaxy, where the magnetic field appears to be mostly perpendicular to the densest matter structures \citep{PIPXXXV2016}. Such perpendicular configuration would produce a negative $T$-$E$ correlation (see Z01). Going from the diffuse ISM to the dense molecular clouds there would be a transition producing more random orientations between the magnetic field and the density structures, reducing the $E-B$ asymmetry. Thus, if the interpretation of the dust polarization power spectra in terms of the correlation between magnetic fields and filamentary density structures is right, one expects a density dependence of the \E-\B\ mode decomposition as well. 

In this paper we present an observational work, in which we extend the \planck\ analysis reported in P16XXX to low Galactic latitude in order to investigate the dependence between the gas column density derived from the {\it Planck} dust emission data and the \E\ and \B\ mode power of dust polarization at $353$~GHz. 
The paper is organized as follows: in Sect.~\ref{sec:data} we describe the \planck\ data used in the analysis; Sect.~\ref{sec:eqs} presents the \E\ and \B\ decomposition and the power spectra at intermediate and low Galactic latitude; in Sect.~\ref{sec:EBNH} we show the correlation between the dust polarization power spectra and the gas column density; in Sect.~\ref{sec:conclusion} we provide the reader with a discussion of our results. A summary is presented in Sect.~\ref{sec:summary}. Two appendices (Appendix~\ref{sec:app} and Appendix~\ref{sec:app1}) clarify our data analysis. 

\section{Data description}\label{sec:data}

In this section we provide a description of the \planck\ polarization data, the column density map, and we describe how we divide the intermediate/low Galactic-latitude sky to define the regions of interest for this analysis.   

\subsection{Planck polarization data}\label{sec:data}

\begin{figure}[!t ] 
   \centering
   \includegraphics[width=0.49\textwidth]{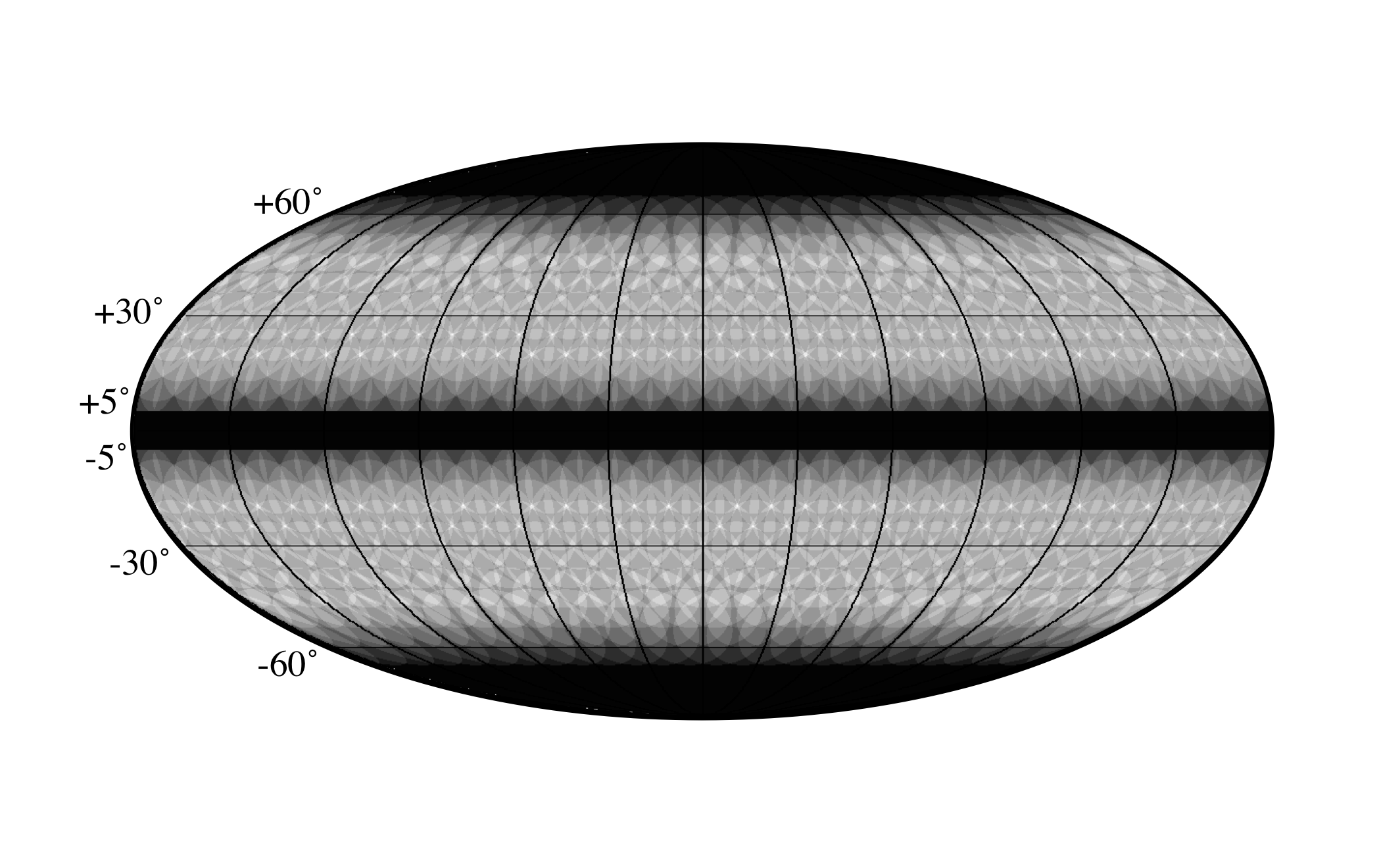}
   \caption{Mask showing the 552 circular sky patches (gray spots) considered in the present analysis. We masked the sky for Galactic latitudes $|b|<5^{\circ}$  and $|b|>60^{\circ}$. The black area is masked. A Galactic coordinates grid centered in ($l,b$)=($0^{\circ},0^{\circ}$) is added with steps of $30^{\circ}$  both in longitude and in latitude.}
\label{fig:hits}
\end{figure}

\begin{figure}[!t] 
   \centering
   \includegraphics[width=10cm]{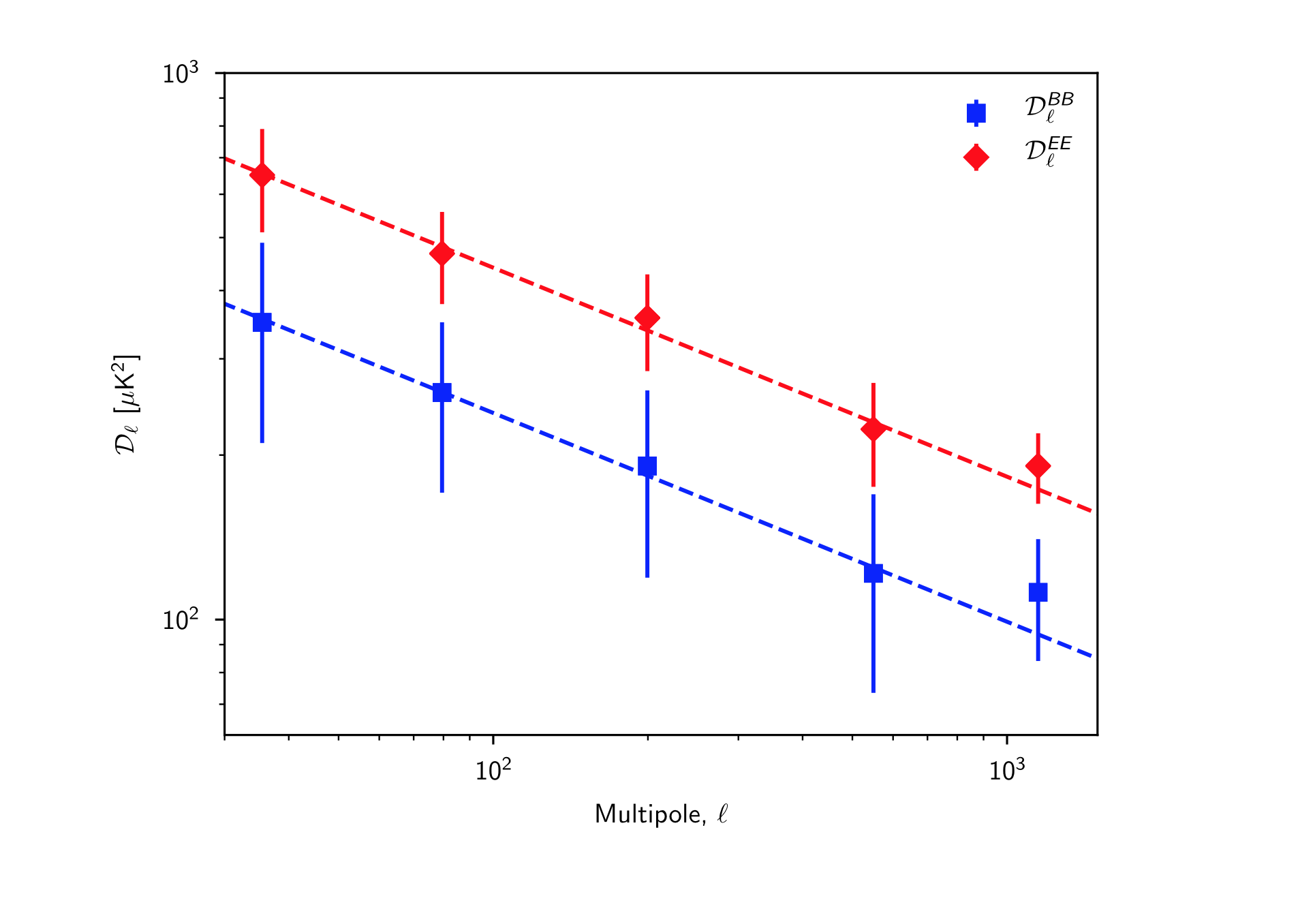}
   \caption{Median values of $\m{D}_{\ell}$ in the five bins in multipole (see text) for \E\ (red) and \B\ (blue) modes over the 552 circular regions in Fig.~\ref{fig:hits}. The dashed lines correspond to the best-fit power-law spectra to the observed Planck data. The slopes of about $-2.4$ are consistent with what presented in P16XXX.}
   \label{fig:EBmodes}
   
\end{figure}
\begin{figure*}[htbp] 
   \centering
   \includegraphics[width=1.\textwidth]{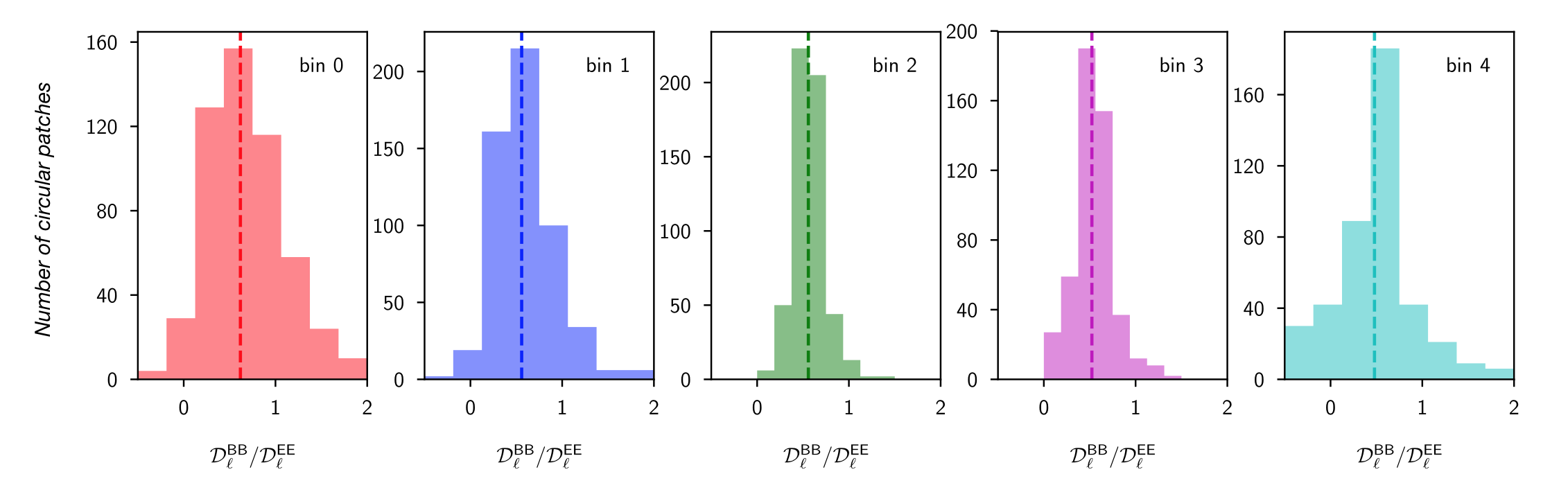}
   \caption{Histograms of the ratio $\m{D}^{BB}_{\ell}/\m{D}^{EE}_{\ell}$ for the multipole bins considered in this work (see text), with centers in: $\ell=35$ (bin 0), $\ell=80$ (bin 1), $\ell=200$ (bin 2), $\ell=550$ (bin 3), $\ell=1150$ (bin 4), with the correspending median values overlaid in dashed vertical lines. In the following analysis we focus on bin 1 to 3.}
   \label{fig:EBmodesRatio}
\end{figure*}

We use publicly available \planck\ PR3 data\footnote{\url{http://www.cosmos.esa.int/web/planck/pla}} at 353 GHz \citep{Plancktest} in \healpix\footnote{\url{http://healpix.sourceforge.net}} format. These maps are produced only
from polarization sensitive bolometers and expressed in thermodynamic temperature units \citep[K$_{\rm CMB}$,][]{Plancktest}. We also use subsets of the \planck\ polarization data at 353 GHz, namely, the half-mission maps (HM1 and HM2), to debias the effect of instrumental noise in the auto-correlation power spectra. We use the raw Stokes $IQU$ maps at 353 GHz at their nominal beam resolution of $4.82\arcmin$ (FWHM).

\subsection{Column density map}\label{ssec:nh}

We consider the total gas column-density map, $N_{\rm H}$, derived from the dust optical depth at $353$~GHz, $\tau_{353}$. The $\tau_{353}$ map \citep{PIPXI2014} was obtained from the all-sky \planck\ intensity observations at $353$, $545$, and $857$~GHz, and the IRAS observations at $100$~$\mu$m, which were fitted using a modified black body spectrum. The $\tau_{353}$ map is used at its nominal resolution of $5\arcmin$. To scale from $\tau_{353}$ to $N_{\rm H}$ we adopted the same convention as in \citet{PIPXXXV2016},
\begin{equation}\label{eq:tau}
\tau_{353}/N_{\rm H} = 1.2 \times 10^{-26}\,{\rm cm}^2.
\end{equation}
Variations in dust opacity are present even in the diffuse ISM
and the opacity increases systematically by a factor of 2 from
the diffuse to the denser ISM \citep{Martin2012,PIPXI2014}. 

In this work, similar to what was done in \citet{PIPXXXV2016}, we want to analyze the column density of local molecular clouds around the Sun. Thus, in order to focus on these dense clouds, and to reduce the contribution to the total $N_{\rm H}$ coming from the large-scale Galactic density gradient, we filter $N_{\rm H}$. The filtered $N_{\rm H}$ map is $\delta N_{\rm H} = N_{\rm H} - \m{N}_{\rm H}$ , where $\m{N}_{\rm H}$ is the column-density map smoothed to a FWHM of $12^{\circ}$. The choice of this scale for the background column density will be clarified in Sect.~\ref{ssec:skydivision}.

As shown in Fig.~\ref{fig:dNH}, the densest regions in $\delta N_{\rm H}$ do correspond to well-known molecular clouds in the Gould Belt: Taurus, Perseus, and California in the extreme East (labeled as A); Cepheus and Polaris in the North-East (labeled as B); Ophiuchus above the Galactic center (labeled as C); Musca and Chamaeleon in the South-West (labeled as D); Orion in the extreme West (labeled as E).

\subsection{Selected sky regions}\label{ssec:skydivision}

In order to study the variations of the \E-\B\ mode power spectra across the sky, we divide it at intermediate and low Galactic latitude ($|b| < 60^{\circ}$) in circular patches of $12^{\circ}$ radius (with an area of $400$\,deg$^2$, or a sky fraction of $f_{\rm sky} \sim 1\%$, see Appendix~\ref{sec:app1}) using a \healpix\ grid at $N_{\rm side}=8$ to get the central pixel of each patch. This radius is chosen to be consistent with the analysis presented in P16XXX. It also explains our choice of filtering $N_{H}$ (see Sect.~\ref{ssec:nh}). To avoid strong depolarization caused by long lines of sight across the Galaxy, we mask the thin Galactic disk for $|b| <5^{\circ}$ \citep{PIPXIX2015}. Hence, we generate a sample of 552 sky patches (see Fig.~\ref{fig:hits}), within which we estimate  average gas column density and dust polarization power spectra.
For each circular patch, the column density value that we consider is represented by the parameter $\delta N^{\star}_{\rm H} =\lan\delta N_{\rm H}(>95\%)\ran$, where the brackets refer to the average over the 5\% densest pixels within each patch. This choice allows us to keep a high dynamic range in column density among the different patches. Results do not significantly change if instead of 5\% we consider 10\%.
\begin{figure*}[ht] 
   \centering
   \includegraphics[width=1.\textwidth]{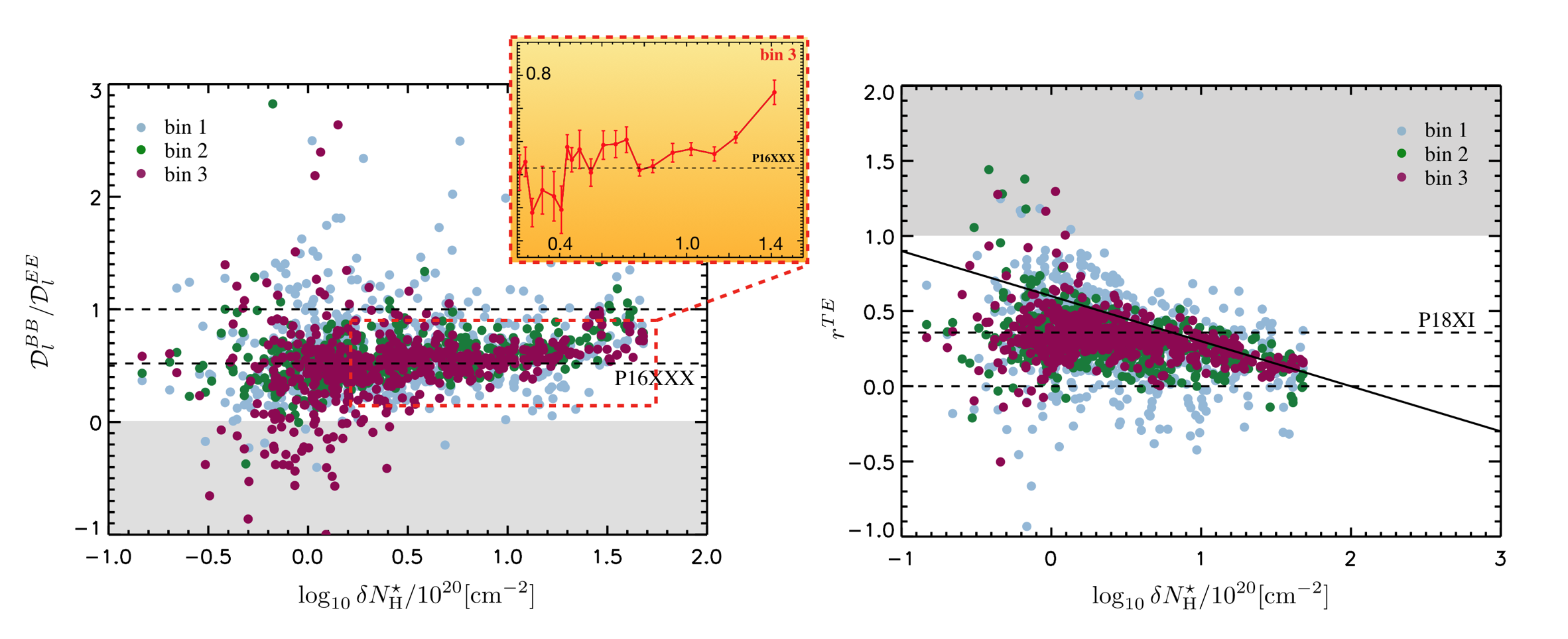}
   \caption{{\it Left panel}: $\m{D}^{BB}_{\ell}/\m{D}_{\ell}^{EE}$ versus the $\delta N^{\star}_{\rm H}$ for all 552 circular regions in bin 1 (light blue), 2 (green), and 3 (purple). Horizontal dashed lines show: i) the value of 1, corresponding to equal power in $E$ and $B$ modes; ii) the value of 0.52, as reported in P16XXX. The inset shows the mean and standard deviation of $\m{D}^{BB}_{\ell}/\m{D}_{\ell}^{EE}$ for bin 3 to highlight the increase of the ratio at high column density. {\it Right panel}: scatter plot of $r^{TE}_\ell$ versus $\delta N^{\star}_{\rm H}$. Colors are same as in the left panel. Horizontal dashed lines show: i) the value of 0, corresponding to absence of correlation between intensity and E modes; ii) the value of 0.36, as reported in P18XI. The solid line shows a fit of $r^{TE}_\ell$ for $\delta N^{\star}_{\rm H} > 10^{21}\,{\rm cm}^{-2}$. The grey-shaded areas in both panels refer to regions dominated by instrumental noise/data systematic effects.}
   \label{fig:EEBBTE}
\end{figure*}
\section{\emph{E-B} mode decomposition: methods}\label{sec:eqs}

This section describes the formalism used to build the \E\ and \B\ mode power spectra from the observed Stokes $Q$ and $U$ parameters. We also show their values within the 552 sky patches introduced in Sect.~\ref{ssec:skydivision}.

\subsection{\emph{E-B} mode formalism}\label{ssec:ebform}
Computing angular power spectra of Stokes parameters requires some discussion. Stokes~$I$ is a scalar quantity that is invariant under rotation. The Stokes~$Q$ and $U$ are not. Following \citet{Zaldarriaga1997} they transform as 
\begin{equation}\label{eq:QUstokes}
(Q +iU)'({{\bm n}}) = e^{\mp 2i\beta}(Q+iU)({{\bm n}}), 
\end{equation}
where ${{\bm n}}$ is the position in the sky and $\beta$ is the rotation of the plane-of-the-sky reference $(\ee_1, \ee_2)$ in $\ee'_1 = \cos{\beta}\, \ee_1 + \sin{\beta}\, \ee_2$ and $\ee'_2 = -\sin{\beta}\,\ee_1 + \cos{\beta}\, \ee_2$. Notice that in the following Stokes~$I$ will be alternatively referred to as $T({{\bm n}})$ for consistency with previous works. The authors of the aforementioned paper expand these quantities in the appropriate spin-weighted basis (spherical harmonics) as 
\begin{align}\label{eq:spharm}
T({{\bm n}}) &= \sum_{\ell m} a_{T,\ell m}Y_{\ell m}({{\bm n}}), \nonumber\\
(Q+iU)({{\bm n}}) &= \sum_{\ell m} a_{2,\ell m} \,\prescript{}{2}Y_{\ell m}({{\bm n}}), \\
(Q-iU)({{\bm n}}) &= \sum_{\ell m} a_{-2,\ell m} \,\prescript{}{-2}Y_{\ell m}({{\bm n}}), \nonumber
\end{align}
and use the spin-raising (lowering) operators, $\eth_+$ ( $\eth_-$ ), in order to get two rotationally-invariant quantities   
\begin{align}\label{eq:spharmrot}
\eth_-^2(Q+iU)({{\bm n}}) &= \sum_{\ell m} \left [ \frac{(\ell+2)!}{(\ell-2)!} \right ]^{1/2} a_{2,\ell m}Y_{\ell m}({{\bm n}}), \\
\eth_+^2(Q-iU)({{\bm n}}) &= \sum_{\ell m} \left [ \frac{(\ell+2)!}{(\ell-2)!} \right ]^{1/2} a_{-2,\ell m}Y_{\ell m}({{\bm n}}). \nonumber
\end{align}
From Eq.~(\ref{eq:spharmrot}), the expansion coefficients are
\begin{align}\label{eq:coeff}
a_{T,\ell m} &=\int Y^{\ast}_{\ell m}({\bm n})T({\bm n}){\rm d}\Omega, \nonumber\\ 
a_{2,\ell m} &= \left [ \frac{(\ell +2)!}{(\ell -2)!} \right ]^{-1/2}\int Y^{\ast}_{\ell m}( {\bm n})\eth_-^2(Q+iU)({{\bm n}}), \\
a_{-2,\ell m} &= \left [ \frac{(\ell +2)!}{(\ell -2)!} \right ]^{-1/2}\int Y^{\ast}_{\ell m}({\bm n})\eth_+^2(Q-iU)({{\bm n}}), \nonumber
\end{align}
which can be linearly combined into
\begin{align}\label{eq:coeffeb}
a_{E,\ell m} &= -(a_{2,\ell m}+a_{-2,\ell m})/2,\\
a_{B,\ell m} &= i(a_{2,\ell m}-a_{-2,\ell m})/2.\nonumber
\end{align}
The $E$ and $B$ modes, scalar and pseudo-scalar fields respectively, are defined as 
\begin{align}\label{eq:eb}
E({{\bm n}}) &= \sum_{\ell m} a_{E,\ell m}Y_{\ell m}({{\bm n}})\\
B({{\bm n}}) &= \sum_{\ell m} a_{B,\ell m}Y_{\ell m}({{\bm n}}).\nonumber
\end{align}
These two quantities are rotationally invariant and they differ for parity symmetries (i.e., changing the sign of the $x$ axis only). Since $Q'({{\bm n}'})=Q({{\bm n}})$ and $U'({{\bm n}'})=-U({{\bm n}})$, from Eqs.~(\ref{eq:coeff}) and (\ref{eq:coeffeb}), one can show that $E'({{\bm n}'})=E({{\bm n}})$ while $B'({{\bm n}'})=-B({{\bm n}})$. Thereby, $E$ and $B$ modes are even and odd quantities, respectively, under parity transformations.

The usual statistical description of the three scalar/pseudo-scalar quantities defined above ($T,E,\,{\rm and}\, B$) is based on their power spectra as a function of the multipole $\ell$,
\begin{equation}\label{eq:cl}
C^{XY}_{\ell} = \frac{1}{2\ell+1}\sum_{m}\langle a^{\ast}_{X,\ell m} a_{Y,\ell m} \rangle,
\end{equation}
where $X$ and $Y$ may refer to $T$, $E$, or $B$. Power spectra are named auto-power spectra when $X=Y$ and cross-power spectra when $X\neq Y$. Alternatively one can use the quantity 
\begin{equation}\label{eq:Dl}
\m{D}^{XY}_{\ell} = \ell(\ell+1)\,{C}^{XY}_\ell/(2\pi).
\end{equation}
In this work we also use the normalized parameter, $r^{XY}$, to quantify the correlation among the power spectra and already shown in P18XI. It is defined as follows,
\begin{equation}\label{eq:rpar}
r^{XY} = \frac{C^{XY}_{\ell}}{\sqrt{C^{XX}_\ell \times C^{YY}_\ell}},
\end{equation}
so that in case of perfect positive (negative) correlation $r^{XY} = 1\,(-1)$, and in case of absence of correlation $r^{XY} = 0$.

\subsection{Power-spectra analysis}\label{ssec:powspec}
We compute the \TEB\ power spectra in Eq.~\ref{eq:cl} for each circular sky-patch using the XPOL\footnote{\url{http://gitlab.in2p3.fr/tristram/Xpol}} code, which is the generalization to polarization of XSPECT \citep{Tristram2005}. XSPECT corrects for incomplete sky coverage, pixel and beam window functions. In order not to correlate noise in the auto-correlated power spectra (i.e, $X=Y$) we always cross-correlate the HM1 and HM2 independent subsets of the data.

We bin the power spectra in five principal multipole-bins centered in $\ell=$35 (hereafter, bin 0), 80 (bin 1), 200 (bin 2), 550 (bin 3), 1150 (bin 4), respectively. The corresponding widths are 15, 40, 200, 500, 1200 from bin 0 to bin 4. 

In Fig.~\ref{fig:EBmodes} we show the median values, and the corresponding standard deviations over the full sample of 552 circular patches of $\m{D}^{EE}_\ell$ (red) and $\m{D}^{BB}_\ell$ (blue) for each selected bin in multipole. On average, the \E-\B\ power spectra at these intermediate/low Galactic latitude are consistent with those presented in P16XXX at high latitude.

The histograms of the $\m{D}^{BB}_{\ell}/\m{D}^{EE}_{\ell}$ ratios for each multipole-bin are displayed in Fig.~\ref{fig:EBmodesRatio}. These distributions enable us to choose a specific selection of bins. In the rest of the analysis we consider neither bin 0 nor bin 4, as bin 0 is highly affected by cosmic variance in small sky patches, and bin 4 is contaminated by noise at full \planck\ resolution (see the corresponding negative tail in $\m{D}^{BB}_{\ell}/\m{D}^{EE}_{\ell}$). As shown in Fig.~\ref{fig:cosmovar}, neglecting bin 0 allows us to ensure, on the other multipole-bins, a level of cosmic variance ($\Delta \m{D}_\ell/\m{D}_\ell$ in the figure) within our 12$^\circ$ circular patches ($\theta_{\rm max}$ in the figure) below 20$\%$.

\section{\emph{E-B} mode power spectra versus $\delta N^{\star}_{\rm H}$}\label{sec:EBNH}

Based on the methodology described above we are now able to study variations of $\m{D}^{BB}_{\ell}/\m{D}^{EE}_{\ell}$, $r^{TE}$, $r^{TB}$, and $r^{EB}$ as a function $\delta N^{\star}_{\rm H}$ for the 552 circular patches at intermediate and low Galactic latitudes.  

In Fig.~\ref{fig:EEBBTE} we show two scatter plots: $\m{D}^{BB}_{\ell}/\m{D}^{EE}_{\ell}$ versus $\delta N^{\star}_{\rm H}$ (left panel) and $r^{TE}$ versus $\delta N^{\star}_{\rm H}$ (right panel). 
In the former one a change in the \E-\B\ asymmetry with column density can be clearly seen. Regardless of the multipole-bin the plot shows that, if for the diffuse ISM, or $\delta N^{\star}_{\rm H} < 3\times 10^{20}$ cm$^{-2}$, $\m{D}^{BB}_{\ell}/\m{D}^{EE}_{\ell}$ is consistent with the value of 0.52 reported in P16XXX, in denser circular patches the ratio tends to increase towards unity; that is, the amount of power in \E\ and \B\ modes for the densest regions is almost the same.

In the right panel of the figure an anti-correlation between $r^{TE}$ and $\delta N^{\star}_{\rm H}$ can be viewed. As in the left panel, $r^{TE}$ is compatible with diffuse ISM value of 0.36 presented in P18XI for $\delta N^{\star}_{\rm H} < 3\times 10^{20}$ cm$^{-2}$. However, as shown by the linear fit of $r^{TE}$ for $\delta N^{\star}_{\rm H} > 10^{21}\,{\rm cm}^{-2}$, for denser regions $r^{TE}$ decreases with column density. The solid line corresponding to the fit could be used to infer the behaviour of $r^{TE}$ if data at higher angular resolution were available. A finer angular resolution would allow one to access to larger column densities otherwise smoothed by the \planck\ beam. This may suggest that $r^{TE}$ would be significantly negative for $\delta N^{\star}_{\rm H} > 10^{22}$ cm$^{-2}$.

Gray shaded areas in both plots define regions where data noise and systematic effects dominate the signal. These are the only causes producing negative values of the $\m{D}^{BB}_{\ell}/\m{D}^{EE}_{\ell}$ and values of $r^{TE}$ larger than unity. We want to stress that the overall scatter of the correlations is not primarily caused by noise, as explained in more detail in Appendix~\ref{sec:app}. It is mostly related to sample variance of a non-Gaussian signal, such as that of interstellar dust polarization, in small sky-patches across the sky. In the same Appendix we also present the 2D probability density function of $\m{D}^{BB}_{\ell}/\m{D}^{EE}_{\ell}$ and $r^{TE}$ (see Fig.~\ref{fig:conts}), which shows an intrinsic anticorrelation between the two parameters. 

\begin{figure}[t!] 
   \centering
   \includegraphics[width=0.5\textwidth]{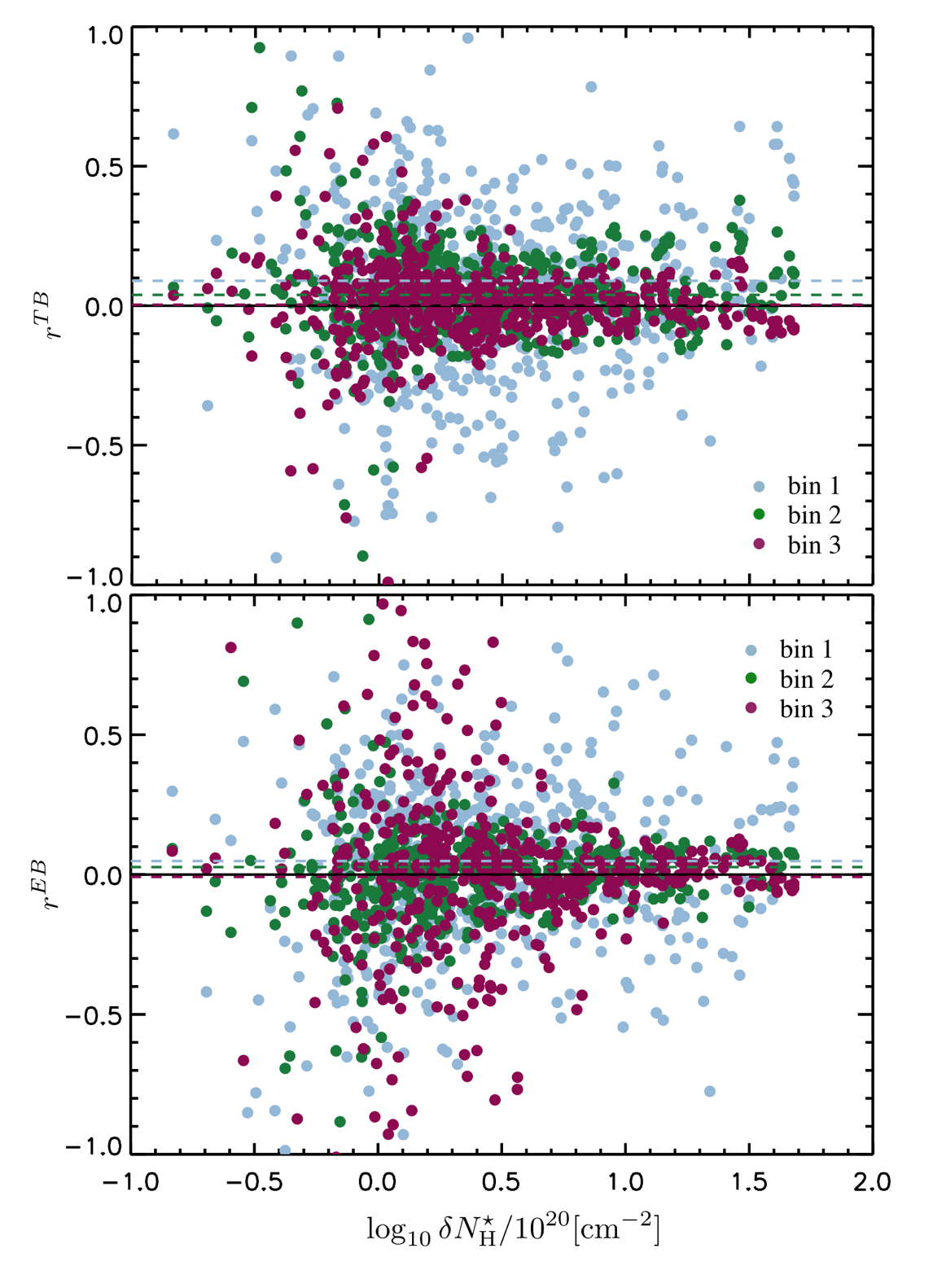}
   \caption{{\it Top panel}: $r^{TB}$ versus $\delta N^{\star}_{\rm H}$. Colors are same as in Fig.~\ref{fig:EEBBTE}. {\it Bottom panel}: $r^{EB}$ versus $\delta N^{\star}_{\rm H}$. Colors are same as in the top panel. Horizontal dashed lines in both panels show the median values of $r^{TB}$ and $r^{EB}$ for $\delta N^{\star}_{\rm H} > 3 \times 10^{20}\,{\rm cm}^{-2}$ in the three multipole bins. Despite the large scatter, a systematic decrease of the median values with angular scale can be seen.}
   \label{fig:TBEB}
\end{figure}

Figure~\ref{fig:TBEB} shows the dependence of $r^{TB}$ and $r^{EB}$ on $\delta N^{\star}_{\rm H}$. These two parameters are noisier. No dependence on column density can be seen. We also find that, in spite of the large scatter, the median values of $r^{TB}$ and $r^{EB}$ for $\delta N^{\star}_{\rm H} > 3\times10^{20}\,{\rm cm}^{-2}$ (see dashed-horizontal lines) are systematically larger, and non-zero, at large scales (bin 1 and bin 2) rather than at small scales (bin 3). As explained in Appendix~\ref{sec:app}, this effect is not due to data noise or to the data analysis.  

\begin{figure*}[t!] 
   \centering
   \includegraphics[width=1.\textwidth]{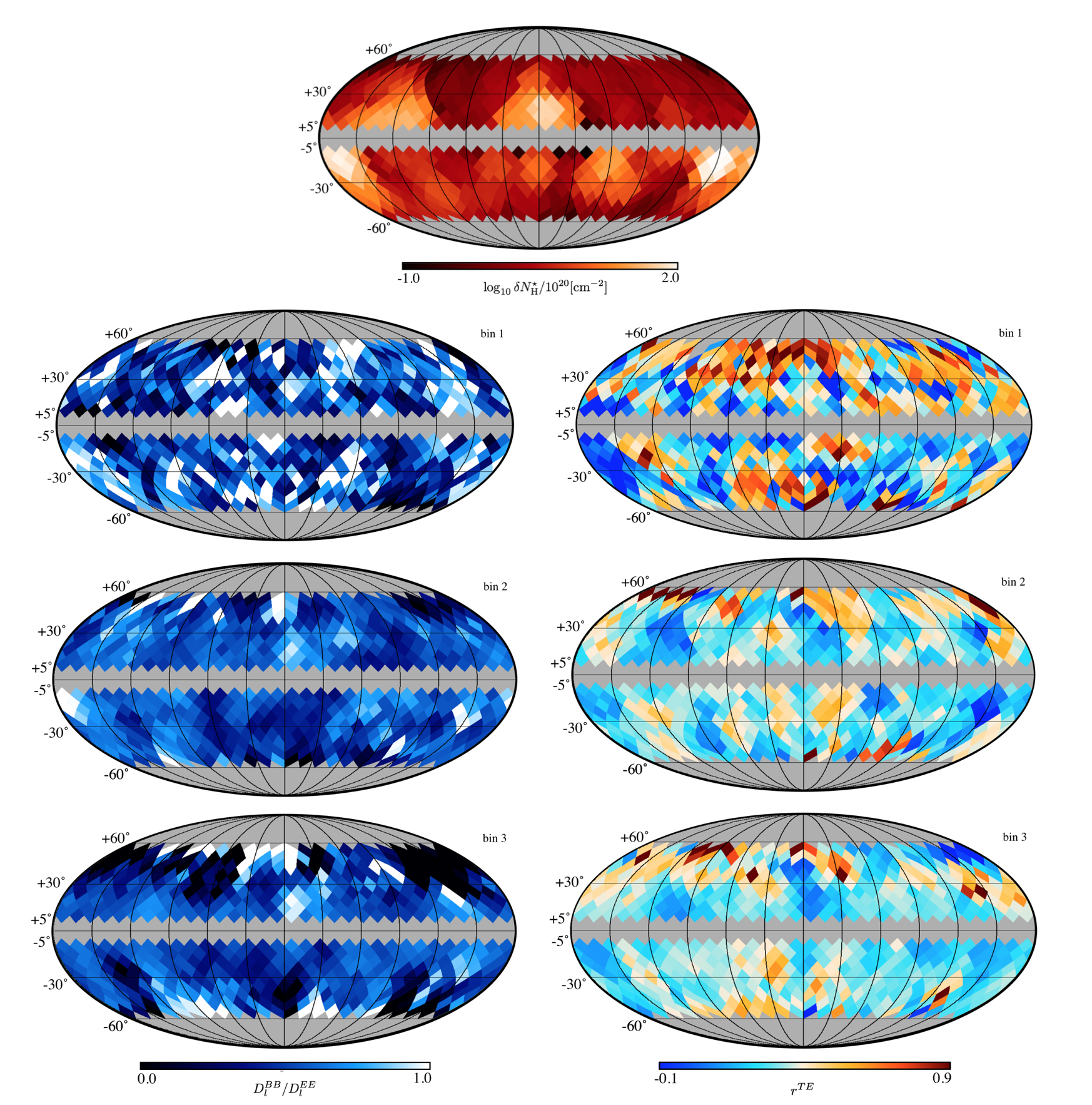}
   \caption{Maps at $N_{\rm side}=8$ of the column-density parameter, $\delta N^{\star}_{\rm H}$ ({\it top-center}), $\m{D}^{BB}_{\ell}/\m{D}_{\ell}^{EE}$ ({\it left}), and $r^{TE}$ ({\it right}) for bin 1, 2 and 3. Non-random patterns changing with multipole, and related to the morphology of $\delta N^{\star}_{\rm H}$, can be seen across the sky.}
   \label{fig:parmaps}
\end{figure*}

For the less noisy parameters, we also produce $N_{\rm side}=8$ maps of  $\m{D}^{BB}_{\ell}/\m{D}^{EE}_{\ell}$ and $r^{TE}$ (see Fig.~\ref{fig:parmaps}). These show how their variations appear correlated with $\delta N^{\star}_{\rm H}$, with organized, non-random, patterns over the intermediate/low latitude sky. 

\section{Discussion}\label{sec:conclusion}
Our work extends the \planck\ analysis on the \E\ and \B\ modes of dust 
polarization at 353 GHz from the diffuse ISM (see P16XXX and P18XI) to denser regions in molecular clouds of the Gould Belt at low Galactic latitude. This study is important both for a better understanding of how interstellar dust affects CMB polarization and for establishing a link between the \E-\B\ 
mode decomposition of dust polarized emission and the ISM physics. 

We focused on the link between the variations of the \E\ and \B\ power spectra, and their 
cross-correlation coefficients ($r^{XY}$, where $X$ and $Y$ are equal to $T$, $E$, or $B$), with $\delta N^{\star}_{\rm H}$ (see Sect.~\ref{ssec:nh}). We confirmed the average values of the \B-to-\E\ power ratio, $\m{D}^{BB}_{\ell}/\m{D}_{\ell}^{EE}$, and $r^{TE}$ in the diffuse ISM ($\delta N^{\star}_{\rm H} < 
3\times 10^{20}$ cm$^{-2}$) reported in P16XXX and P18XI. However, for denser regions ($\delta N^{\star}_{\rm H} > 
3\times 10^{20}$ cm$^{-2}$) we found clear departures from these mean values with signs of {\it correlation} between $\m{D}^{BB}_{\ell}/\m{D}_{\ell}^{EE}$ and $\delta N^{\star}_{\rm
H}$, and of {\it anticorrelation} between $r^{TE}$ and $\delta N^{\star}_{\rm H}$. We found as well an intrinsic anticorrelation between $\m{D}^{BB}_{\ell}/\m{D}_{\ell}^{EE}$ and $r^{TE}$. 

These results strengthen the interpretation of the \E-\B\ asymmetry, and the positive $TE$ correlation, in terms of the alignment between the magnetic-field orientation and the density filamentary structures in the ISM, as already claimed in \citet{planck2015-XXXVIII} for the diffuse medium. A positive $r^{TE}$ and a $\m{D}^{BB}_{\ell}/\m{D}_{\ell}^{EE}$ less then unity, would be both naturally produced by filamentary structures aligned with the orientation of the interstellar magnetic field (see Z01), which was proved true in the diffuse ISM at high latitude by \citet{PIPXXXII} and \citet{planck2015-XXXVIII}.

The same alignment was also observed in the diffuse surrounding of molecular clouds in the Gould Belt. However, these lower-latitude regions present as well a gradual change in relative orientation, or a smooth transition from parallel to perpendicular, for denser and denser matter structures with respect to the magnetic field \citep{PIPXXXV2016,Soler2017}. This change in relative orientation is considered representative of the dynamical properties of molecular clouds. Based on the comparison between data and numerical simulations, the change in relative orientation with increasing matter density is indicative of molecular clouds dominated by their self-gravity in sub/trans-Alfv\'enic MHD turbulent media \citep{Soler2013}. Always following Z01, a relative perpendicular orientation between filamentary density structures and the magnetic field would still produce $\m{D}^{BB}_{\ell}/\m{D}_{\ell}^{EE} < 1$ but values of $r^{TE} < 0$. The extrapolation of $r^{TE}$ with $\delta N^{\star}_{\rm H}$ in the right panel of Fig.~\ref{fig:EEBBTE} indeed shows that, for $\delta N^{\star}_{\rm H} > 10^{22}$ cm$^{-2}$, $r^{TE}$ may gradually change and become negative. This value of column density is also very close to that quoted in \citet{PIPXXXV2016} that corresponds to the change in relative orientation. The smooth change in relative orientation between density filamentary structures and magnetic-field orientation would produce a transition in the values of $\m{D}^{BB}_{\ell}/\m{D}_{\ell}^{EE}$, which, as shown in the left panel of Fig.~\ref{fig:EEBBTE}, would first increase towards unity and decrease again once most of the dense structures would be perpendicular to the magnetic-field orientation. However, at the angular resolution of \planck\, we do not have access to enough statistics for tracing the densest filamentary structures in molecular clouds \citep{PIPXXXIII}. Thus, in order to confirm this interpretation, higher angular resolution polarization surveys to probing interstellar dust emission would be necessary \citep[e.g., BFORE,][]{Bryan2018}.

Another result that extends the recent finding of P18XI is that $r^{TB}$, and maybe 
$r^{EB}$, may indeed differ from zero, with a stronger positive signal at large scale compared to small scales. \citet{Bracco2019} suggested that the $TB$ positive correlation at very large scale (for multipoles $l<50$) may be principally caused by the Galactic-magnetic field structure in the Solar neighborhood, which would leave an imprint of a left-handed helical component on the $TB$ correlation on scales of a few hundred parsecs. However, at the angular scales probed in this work, other processes may be at play, since for the closest Gould Belt clouds we would be probing physical scales of a few parsecs. Further investigation is needed to understand what kind of mechanisms may generate the $TB$ correlation in molecular clouds.

From previous works it is worth to notice that most of the effort was put to understand the level of \E-\B\ asymmetry. Our analysis shows that, although such value is on average true in the diffuse ISM, large variations are found across the sky. These variations have organized patterns at intermediate and low Galactic latitude (Fig.~\ref{fig:parmaps}). They must be related to intrinsic changes in ISM physics and interstellar dust properties. 

\citet{Kim2019} used MHD simulations to produce all-sky synthetic observations to study the \E-\B\ asymmetry. They concluded that the observed power spectra strongly fluctuate depending both on the position of the observer and on temporal fluctuations of ISM properties due to variations of the star formation process. For the first time, our work shows that the level of \E-\B\ asymmetry in real observational data may indeed significantly vary depending on the sky position. However, comparing observational data and all-sky non-Gaussian stochastic models of dust polarization, we showed that most of the variations of \E-\B\ modes in the diffuse ISM are likely due to sample variance across the sky rather than to intrinsic physical differences among the sky patches. This is not true in the dense ISM, where the \E-\B\ decomposition depends on the value of the gas column density, thus likely on the physics of the observed ISM region. This is important for modelling the impact of dust polarization in CMB studies and for assessing the link between \E-\B\ modes and ISM physics.  

\section{Summary}\label{sec:summary}

We have presented a novel analysis of the \planck\ polarization data at 353 GHz that extends the study of the $T$-\E-\B\ mode power spectra of interstellar dust to low Galactic latitude ($|b| < 60^{\circ}$ and $|b| > 5^{\circ}$). We investigated the correlation between these power spectra and the gas column density, which, in the selected sky, is dominated by the emission of  molecular clouds in the Gould Belt. Our analysis is relevant to better characterize the statistical properties of dust polarization, both to model Galactic foreground emission to the CMB polarization and to study the dynamical properties of the ISM.

We divided the selected sky in 552 identical circular patches within which we could define mean values of column density, $\delta N^{\star}_{\rm H}$, and of $T$-\E-\B\ power spectra for multipoles between $80<\ell<550$. We thus studied the respective auto and cross correlations ($r^{XY}$, with $X$ and $Y$ equal to $T,\, E, \, B$). The main results of our work are listed in the following:
\begin{itemize}
    \item we found that the \B-to-\E\ power ratio, $\m{D}^{BB}_{\ell}/\m{D}_{\ell}^{EE}$, correlates with column density, $\delta N^{\star}_{\rm H}$. While for $\delta N^{\star}_{\rm H} < 3\times 10^{20}$ cm$^{-2}$ the values of $\m{D}^{BB}_{\ell}/\m{D}_{\ell}^{EE}$ are consistent with what was already found in the diffuse ISM ($\m{D}^{BB}_{\ell}/\m{D}_{\ell}^{EE} \approx 0.5$, P16XXX, P18XI), for larger column density the ratio increases approaching unity; 
    \item we found that the positive $TE$ correlation observed in the diffuse ISM ($r^{TE}\approx0.36$, P18XI) is on average compatible with our results for $\delta N^{\star}_{\rm H} < 3\times 10^{20}$ cm$^{-2}$. However, for denser regions we found a clear anticorrelation between $r^{TE}$ and $\delta N^{\star}_{\rm H}$, with $r^{TE}$ approaching zero for our densest sample of column density in molecular clouds of the Gould Belt. This trend suggests that $r^{TE}$ could become negative for $\delta N^{\star}_{\rm H} > 10^{22}$ cm$^{-2}$, corresponding to a perpendicular relative orientation between density structures and magnetic field in molecular clouds (see Z01). This would be consistent with the analysis of histograms of relative orientations in dense molecular clouds \citep[i.e.,][]{PIPXXXV2016}. Only high-resolution polarization surveys of dust emission will allow us to confirm this interpretation;  
    \item we found an anticorrelation between $r^{TE}$ and $\m{D}^{BB}_{\ell}/\m{D}_{\ell}^{EE}$;
    \item we confirmed that, as shown in P18XI, the median value of the $r^{TB}$ may be positive and non-zero at large scale (for multipoles $l \approx 80$). We did not find any dependence between $\delta N^{\star}_{\rm H}$ and $r^{TB}$, or  $r^{EB}$, however, this may be due to the low signal-to-noise in $TB$ and $EB$;
    \item we found that the \E-\B\ mode dust power spectra show strong variations compared to the mean values reported in previous works. These variations, seen correlated on the sky, are not due to noise. In the diffuse ISM they are mainly caused by small sample variance of a highly non-Gaussian signal such as interstellar dust polarization. In the dense ISM, however, they appear correlated with the column density suggesting that we may effectively trace changes of ISM physical properties (i.e., Galactic magnetic field structure, interstellar turbulence). This is both relevant to model the impact of dust polarization as a CMB foreground and for understanding the link between the \E-\B\ mode decomposition and ISM physics.
\end{itemize}

\begin{acknowledgements}
We gratefully acknowledge the use of the Aquila cluster at NISER, Bhubaneswar. This research is partly supported by the Agence Nationale de la Recherche (project BxB: ANR-17-CE31-0022). Some of the results in this paper have been derived using the \healpix\ \citep{Gorski2005} package. 
\end{acknowledgements}

\bibliographystyle{aa}
\bibliography{aanda.bbl}

\appendix 
\section{Non-Gaussian simulations of dust polarization}\label{sec:app}
\begin{figure*}[!h] 
   \centering
   \includegraphics[width=1.\textwidth]{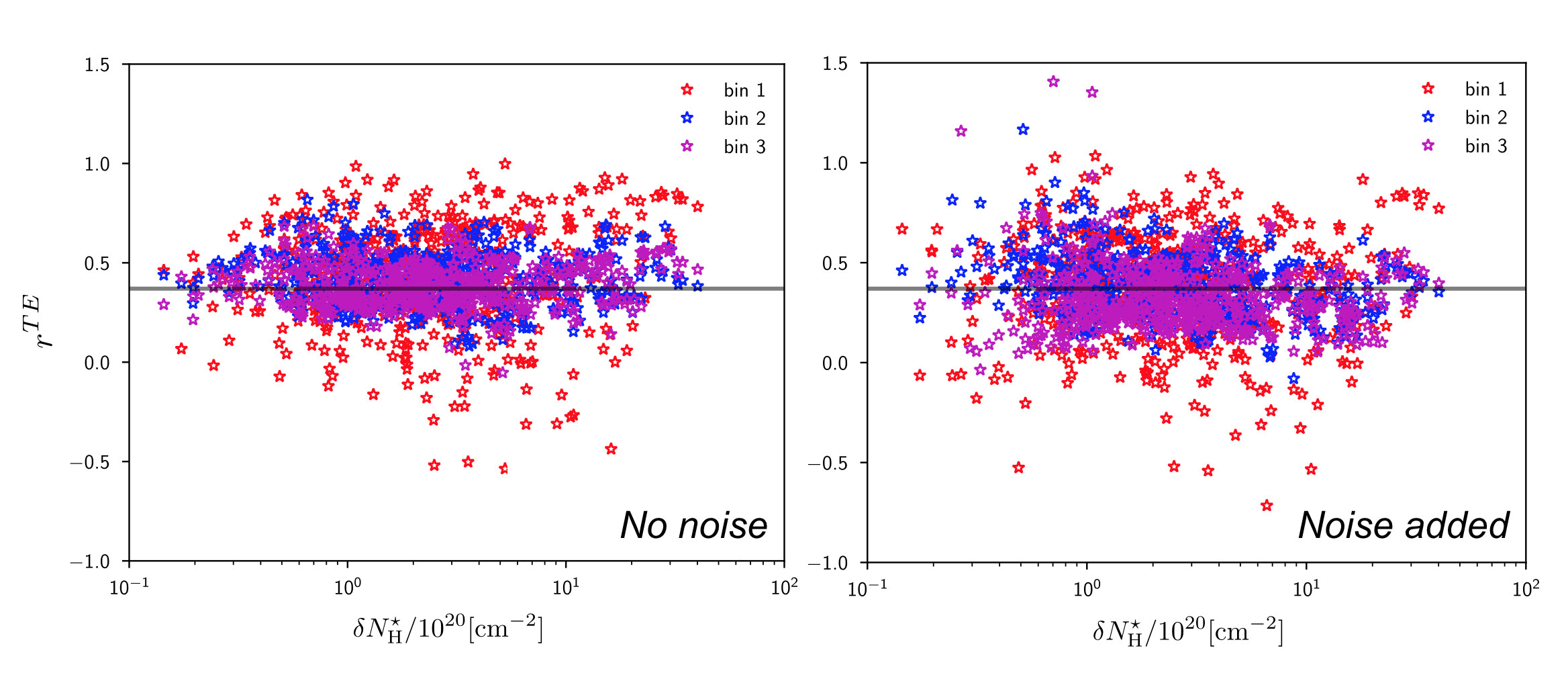}
   \caption{Scatter plots of modeled $r^{TE}$ obtained with the simulations against the gas column density derived from the data, $\delta N^{\star}_{\rm H}$, with ({\it right panel}) and without ({\it left panel}) \planck\ noise. No dependence with $\delta N^{\star}_{\rm H}$ is expected. The variance observed in the scatter plots is not dominated by noise. Notice that the median values in every multipole bin (see horizontal solid lines) correspond to the input values used to generate the simulations.}
   \label{fig:GaussTest}
\end{figure*}
\begin{figure*}[ht] 
   \centering
   \includegraphics[width=1.\textwidth]{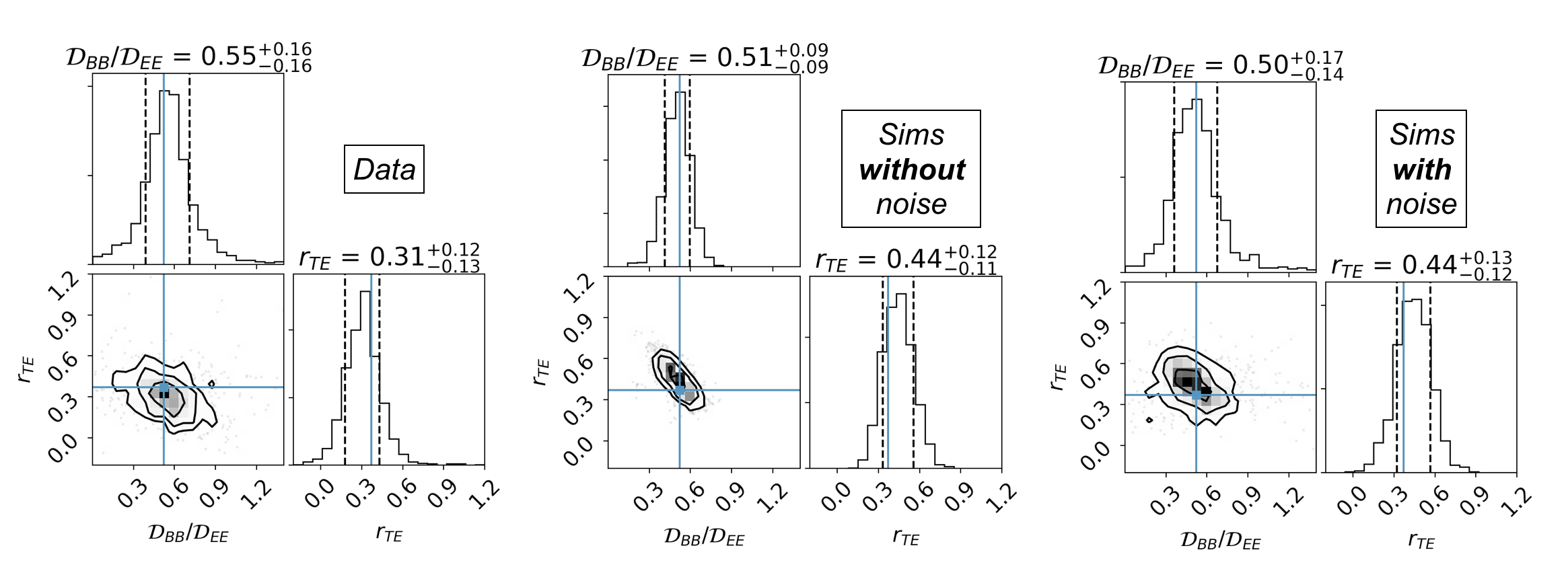}
   \caption{Probability density functions (pdf, 1D and 2D) of $r^{TE}$ and $\m{D}^{BB}_{\ell}/\m{D}_{\ell}^{EE}$ only for bin 2 and bin 3 together with 1-, 2-, 3-$\sigma$ contour levels for the observed data ({\it left panel}), the simulations without \planck\ noise ({\it central panel}), and the simulations with \planck\ noise added ({\it right panel}). The blue lines indicate the observed values in P16XXX and P18XI of $\m{D}^{BB}_{\ell}/\m{D}_{\ell}^{EE} = 0.52$ and $r^{TE} = 0.36$. On top of the 1D pdf the mean values and the standard deviations (also represented by black vertical dashed lines) are shown. These plots were made with the PYTHON CORNER package \citep{corner}.}
   \label{fig:conts}
\end{figure*}
In this Appendix we test the methodology described in Sect.~\ref{ssec:powspec}, performing the same analysis on non-Gaussian simulations of the polarized sky, which have the property of reproducing the 1- and 2-point statistics of the \planck\ polarization data at high Galactic latitude \citep{PIPXLIV2016, Vansyngel2017}. These simulations are stochastic models of polarized dust emission on the sphere. The method builds on the understanding of Galactic polarization in terms of the structure of the Galactic magnetic field and its coupling with interstellar matter and turbulence through a handful of parameters. The simulated maps do not correspond to Gaussian random fields as shown in Fig.~5 of \citet{Vansyngel2017}. 

We generate two sets of simulations, with and without a noise realization from {\planck} (including systematic effects), in which the input values of $r^{TE}$ and $\m{D}^{BB}_{\ell}/\m{D}_{\ell}^{EE}$ are fixed to 0.36 and 0.52, respectively. Results can be seen in Fig.~\ref{fig:GaussTest} for the simulated $r^{TE}$ parameter against the observed $\delta N^{\star}_{\rm H}$. The effect of noise does not increase significantly the variance in the correlation plots, which is dominated by the intrinsic variance among the different sky patches in the simulations, as will be detailed in the following. Moreover, as expected, no dependence exists between the observed $\delta N^{\star}_{\rm H}$ and the simulated $r^{TE}$. We also notice that, regardless of the multipole bin, the input values for the medians of $r^{TE}$ (solid horizontal lines) are obtained in output. The same is found for the simulated $r^{TB}$ and $r^{EB}$ parameters where the input values are set to 0. These two parameters do not show any systematic decrease in the median values with scale as observed in the \planck\ data. Thus, we conclude that the decrease in the median values of $r^{TB}$ and $r^{EB}$ observed in the data, from large to small scale, cannot be caused by noise or by our methodology. We suggest that, unless residual (unknown) systematic effects in the data are present, the observed decrease may be true. However, due to the large scatter in the distributions of the observed $r^{TB}$ and $r^{EB}$, it is not possible to quantify the significance of this statement.

In Fig.~\ref{fig:conts} we show 1D and 2D probability density functions for $r^{TE}$ and $\m{D}^{BB}_{\ell}/\m{D}_{\ell}^{EE}$ for the observed and the simulated data, respectively, considering bin 2 and bin 3 together in order to increase the number statistics. The two parameters appear clearly anticorrelated both in the observations and in the simulations. From Eq.~5 in \citet{Vansyngel2017} the inverse dependence in the simulated spectra can be derived as $\m{D}^{BB}_{\ell}/\m{D}_{\ell}^{EE} \approx -m(r^{TE})^2 + q$, where $m = 0.54$ and $q = 0.56$. 
The effect of noise smooths the anticorrelation between $r^{TE}$ and $\m{D}^{BB}_{\ell}/\m{D}_{\ell}^{EE}$, suggesting that the true anticorrelation in the \planck\ data is likely stronger. The spread about the mean values found in the noisy simulations allows us to statistically recover the observed data dispersion (see the standard deviations quoted in the figure), confirming that sample variance is a major responsible for the \TEB\ power fluctuations from patch to patch over the sky at least in the diffuse ISM. This result validates the simulations presented in \citet{Vansyngel2017} for the statistical description of the polarized properties of the diffuse ISM even in small sky patches. However, as proved by our work, a significant dependence of the parameters with column density for $\delta N_{\rm H}^{\star} > 3\times 10^{20}$ cm$^{-2}$ is observed. This is not captured yet by any existing model.

\section{Cosmic variance per multiple bin}\label{sec:app1}

\begin{figure}[h!]
   \centering
   \includegraphics[width=.5\textwidth]{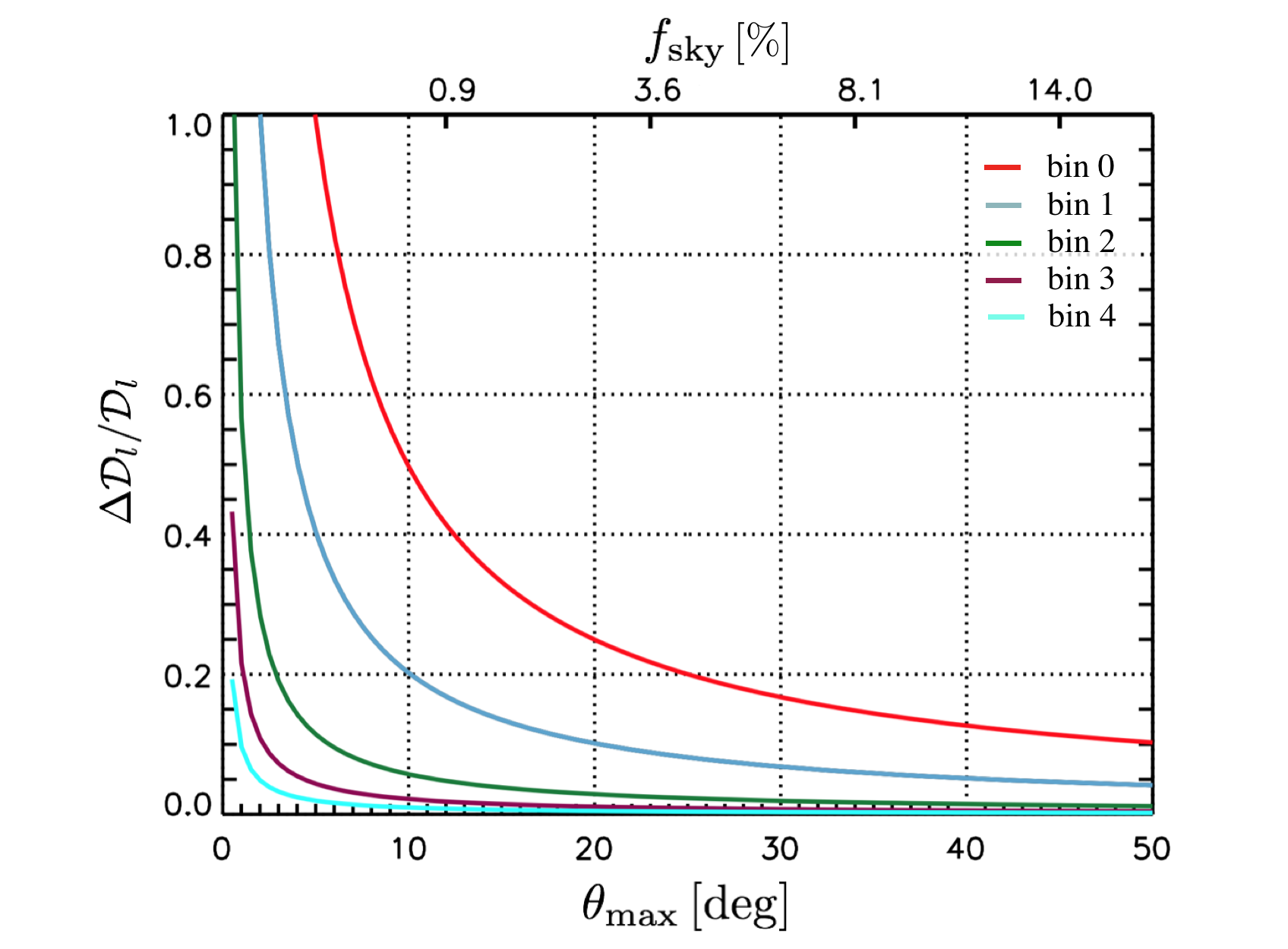}
   \caption{Cosmic variance per multipole bin versus sky-patch size and sky-fraction \citep{Tegmark1997}. The central multipoles for each bin are: $\ell=35$ (bin 0), $\ell=80$ (bin 1), $\ell=200$ (bin 2), $\ell=550$ (bin 3), $\ell=1150$ (bin 4).}\label{fig:cosmovar} 
\end{figure}

We show a figure that allows us to quantify the level of cosmic variance in each multipole bin used in the data analysis. Following \citet{Tegmark1997}, the cosmic variance can be estimated as
\begin{equation}
    \frac{\Delta{\m{D}_\ell}}{\m{D}_\ell} \approx \sqrt{\frac{2}{(2\ell+1) L f_{\rm sky}}},
\end{equation}
where $f_{\rm sky}$ is the sky fraction considered and related to the sky-patch size as $f_{\rm sky} = \sin^2{(\theta_{\rm max}/2)}$, which in our case is $\theta_{\rm max}=12^{\circ}$; and $L$ is the width of the $\ell$-bin equal to 15, 40, 200, 500, and 1200 from bin 0 to bin 4 respectively. As shown in Fig.~\ref{fig:cosmovar}, neglecting bin 0 enables us to limit the level of cosmic variance per bin below 20\%. Notice that this equation is not completely accurate for cross-spectra. In that case it would read $\Delta D_\ell^{XY}=(1/\nu_\ell)[(D_\ell^{XY})+D_\ell^{XX}D_\ell^{YY}]$. Moreover, this estimates are only valid in case of Gaussian random fields. The observed signal is not Gaussian, thus we expect a larger amount of variance per bin of a factor of a few \citep{Vansyngel2017}.

\end{document}